Hybrid Materials

# Formation and Thermal Stability of Gold–Silica Nanohybrids: Insight into the Mechanism and Morphology by Electron Tomography**

Paromita Kundu,* Hamed Heidari, Sara Bals, N. Ravishankar, and Gustaaf Van Tendeloo

**Abstract**: Gold–silica hybrids are appealing in different fields of applications like catalysis, sensorics, drug delivery, and biotechnology. In most cases, the morphology and distribution of the heterounits play significant roles in their functional behavior. Methods of synthesizing these hybrids, with variable ordering of the heterounits, are replete; however, a complete characterization in three dimensions could not be achieved yet. A simple route to the synthesis of Au-decorated $SiO_2$ spheres is demonstrated and a study on the 3D ordering of the heterounits by scanning transmission electron microscopy (STEM) tomography is presented—at the final stage, intermediate stages of formation, and after heating the hybrid. The final hybrid evolves from a soft self-assembled structure of Au nanoparticles. The hybrid shows good thermal stability up to 400 °C, beyond which the Au particles start migrating inside the $SiO_2$ matrix. This study provides an insight in the formation mechanism and thermal stability of the structures which are crucial factors for designing and applying such hybrids in fields of catalysis and biotechnology. As the method is general, it can be applied to make similar hybrids based on $SiO_2$ by tuning the reaction chemistry as needed.

Gold-based heterostructures are important in different fields of applications starting from catalysis to biomedicine.[1–4] A good control over morphology, stability, and ordering of the heterounits is a prime requirement for most applications. For example, in catalytic applications, reactions are often carried out at higher temperatures and retaining the size and dispersion of the catalyst particles on the support material is critical as the active surface area inversely scales with the size.[5] In this context, $SiO_2$-based metal hybrids are desired because they are readily synthesized, show good thermal and chemical stability, have tunable optical properties, which are size-dependent, and benefit from the inertness of $SiO_2$ as a support/coating material. Different strategies exist to obtain these hybrids, with various kinds of distribution and morphology, using chemical routes[6–9] but often the attachment of nanoparticles on the $SiO_2$ sphere is not favorable because of the high interfacial energy or negatively charged silica surface and hence functionalization of $SiO_2$ is employed to modify the interfacial energies. For instance, Au nanoparticles do not nucleate favorably on a $SiO_2$ surface without functionalization.[7,10] However, particles attached to $SiO_2$ being held by a monolayer of linkers on the surface are often less stable because of the high mobility of the Au nanoparticles on $SiO_2$, as well as the low thermal stability of the molecular chains. Therefore, new strategies are required to design more stable hybrid structures. Also, it is necessary to understand the mechanism of formation and the morphological changes that such structures might undergo in a course of reaction as a function of temperature. In heterostructures, the ordering of the component units often determines their functionality. It is therefore important to determine the distribution of the catalyst nanoparticles on the support materials with different morphologies. For such heterostructures, conventional imaging techniques such as TEM and STEM may give information on the size of the nanostructures. However, it is important to realize that these techniques yield 2D projection images, which can render misleading information concerning the shape, morphology, and distribution of the heterounits. For instance, metal nanoparticles coated with a silica layer and vice versa have been reported but the uniformity or continuity of the coating cannot be confirmed from 2D projection images only. Hence, the actual morphology of the nanostructures often remains elusive without a 3D characterization. Electron tomography is a state-of-the-art technique to perform such an investigation. A 3D reconstruction of an object is obtained from a series of its 2D projection images and can give precise information on the shape, position and distribution of the heterounits.[11,12]

In this report, we present a facile synthetic route to Au–$SiO_2$ spheres in which the Au nanoparticles are decorated in the outermost layer of the formed silica spheres. Formation of the hybrid is mediated by self-assembling of Au nanoparticles capped with a mixture of oleyl amine and MPTMS. High-angle annular dark-field (HAADF) STEM tomography and X-ray energy dispersive spectroscopy (XEDS) have been performed to investigate the structure and composition of the hybrid. A mechanism has been proposed to understand the formation of such hybrids in the medium. The thermal stability of the hybrid has been investigated that indicates a low mobility of the Au nanoparticles in the hybrid. However, penetration of the particles from the surface into the matrix of the silica sphere was observed from the electron tomography experiments. This study provides useful insights in the morphological changes that may be encountered by similar hybrids in the course of temperature-dependent processes such as catalysis.

[*] Dr. P. Kundu, Dr. H. Heidari, Prof. S. Bals, Prof. Dr. G. Van Tendeloo
Electron Microscopy for Materials Research (EMAT)
University of Antwerp
Groenenborgerlaan 171, 2020 Antwerp (Belgium)
E-mail: paromita.kundu@uantwerp.be
Prof. N. Ravishankar
Materials Research Center, Indian Institute of Science
C.V. Raman Avenue, Bangalore, 60012 (India)

[**] This research has received funding from the European Community's Seventh Framework Program (ERC; grant number 246791)—COUNTATOMS, COLOURATOMS, as well as from the IAP 7/05 Programme initiated by the Belgian Science Policy Office. Funding from the Department of Science and Technology (DST) is also acknowledged.

Gold nanoparticles of 2–10 nm size are formed by a microwave reduction process using oleyl amine as the reducing and capping agent. After treating these particles with a silica precursor, MPTMS, and sodium silicate in ethanol medium, an Au-SiO$_2$ hybrid formed, which is composed of 300-500 nm sized spheres with a silica core and Au nanoparticles assembled on its surface as shown in Figure 1 a,b. The high-resolution image (inset of b) shows an image of a faceted Au particle with the {111} and {001} planes proximity to the Au nanoparticles implies that the oleyl amine capping exists along with the thiol but it remains mostly on the Au surface, outside the silica matrix. This is also evident from the line scan profile (see section S1). To reconstruct the 3D shape of the hybrids, HAADF-STEM tomography was performed and HAADF-STEM images were acquired by tilting the specimen from +$f$ to -$f$ degrees (where $f$ varies for different samples and is typically within ±70°) with increments of 2°. Examples of 2D projection images are presented in Figure 2 a. The Au particles appear with higher intensity because of the chemical sensitivity of the HAADF-STEM technique. 3D reconstructions were obtained using 30 iterations of the SIRT algorithm.[5] The visualizations, presented in Figure 2 b, show that the fully formed spheres yield Au nanoparticles only at the surface and not embedded in the matrix. This is especially obvious from the orthoslices presented in Figure 2 c,d, where the Au particles are clearly

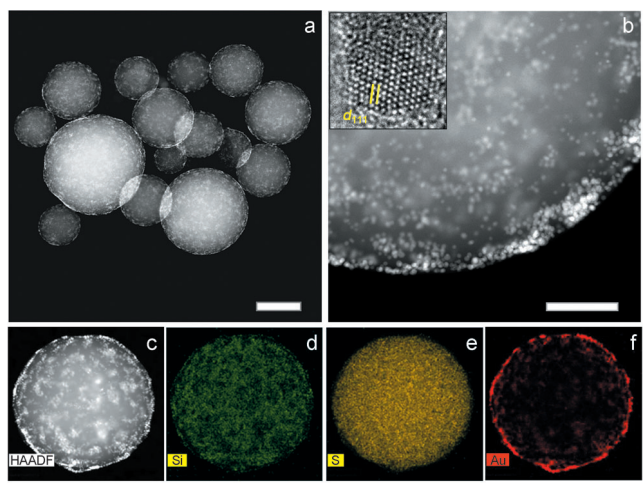

*Figure 1.* a and b) Low- and high-magnification HAADF-STEM images of an as-synthesized hybrid, respectively, show the Au particles and their average distribution on the SiO$_2$ spheres (the scale bars are 500 and 50 nm, respectively) and the inset of (b) shows the high-resolution transmission electron microscopic image (HRTEM) from an Au particle, resolving the {111} and {100} planes. c–f) HAADF images of a selected hybrid particle with the elemental maps for Si-K, S-K, and Au-L lines.

resolved. The size of the Au particles was retained in the hybrid although they were loosely clustered at several places. HAADF-STEM images, presented in Figure 1 a,b show the distribution of the Au nanoparticles on the SiO$_2$ spheres; however, it is difficult from such 2D images to determine the location of the Au particles on the silica. It is impossible to conclude if they are present only at the surface or embedded into the matrix of the silica spheres. STEM-EDX elemental mapping is performed to confirm the composition of the hybrid (EDX = energy-dispersive X-ray; Figure 1 c–f). Figure 1 f shows the Au-L map which indicates a large density of Au nanoparticles on the periphery in comparison to the center. The Si-K and S-K maps are given in Figure 1 d and e, respectively, and the O-K map closely overlaps with that of Si (see section S1 in the Supporting Information); it is therefore clear that SiO$_2$ along with S-containing MPTMS forms the dense three-dimensional mass on which the Au nanoparticles are attached presumably by -SH (thiol) moieties of MPTMS. However, the presence of N (N-K map in section S1) in close

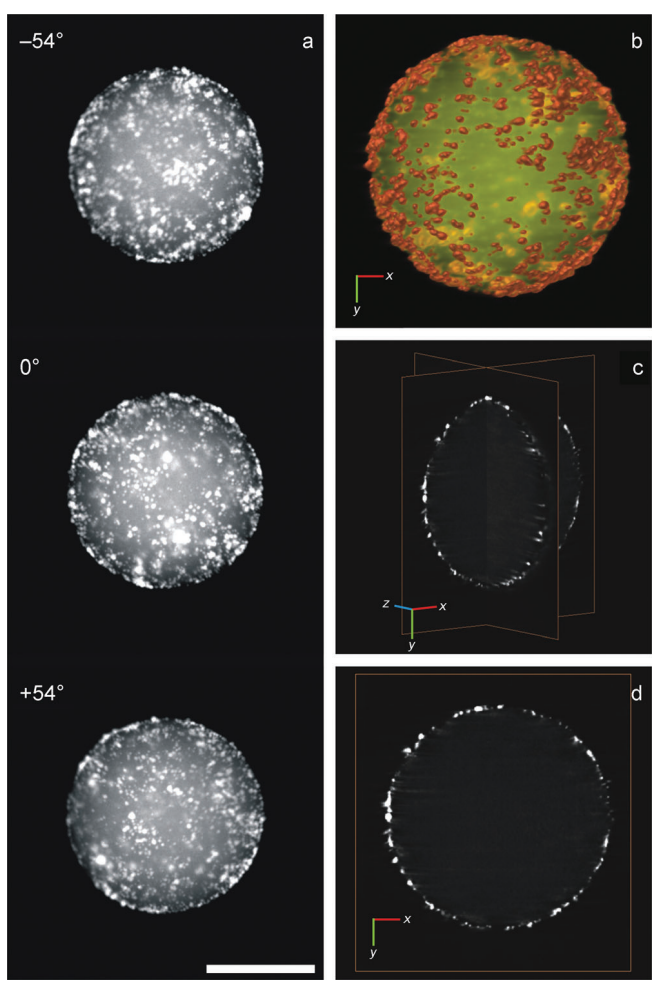

*Figure 2.* STEM tomography of an Au-SiO$_2$ particle. a) 2D projection images at 0° and higher (±54°) tilt angles (the scale bar is 200 nm). b) 3D volume rendering obtained from the reconstruction of the aligned tilt series images, recorded by tilting the sample from −70° to +56° with 2° increment, showing that the Au particles are located on the surface of the sphere. c) Intersection of two different orthoslices. d) The orthoslice in the *xy*-plane shows the presence of Au particles, with brighter Z contrast, only at the surface of the sphere and the SiO$_2$ matrix inside corresponds to lighter Z contrast.

present only at the surface of the SiO$_2$ matrix; a 3D animation is presented in section S2.

Formation of Au-SiO$_2$ hybrids with the observed morphology and composition has never been investigated earlier. However, understanding the formation mechanism could provide useful insight to tune the reaction parameters and conditions in order to exercise control over size, shape, and distribution. Previous reports describe competitive reaction schemes for the thiol-containing silane precursors to undergo hydrolysis, polymerization, and monolayer formation on Au.[13] As these reactions are largely dependent on the reactant concentration, it is difficult to resolve the steps sequentially towards the formation of the hybrids discussed here. However, based on the investigation of the intermediate structures formed during the reaction, we propose a formation mechanism which can lead to the Au-SiO$_2$ structures. Figure 3a–d shows different types of Au nanoparticle assemblies that resulted after 4–5 h of reaction and aggregates of formed silica spheres, some of them being already coated with Au nanoparticles, were also found. Of these types, the self-assembly as shown in Figure 3a, was found to be present in large majorities during the reaction. Figure 3e shows a HAADF-STEM image of the assembled Au particles. The STEM-EDX elemental maps in Figure 3(f–h) present evidence of binding of the S- and N-containing ligands to the Au particles. The Si-K and O-K maps (see section S4) reveal the absence of a SiO$_2$ core and that the assemblies are hollow. The formation of all these products and the final Au-SiO$_2$ morphology can be understood considering the binding of the two types of ligands to the Au particles, their interaction with the solvent medium and the chemistry involved in the formation of the silica spheres. In Figure 3i we describe schematically the stages of formation of the Au-SiO$_2$ hybrid in solution (an elaborate reaction scheme is shown in section S3). An aliquot of a particular concentration of MPTMS in ethanol (polar medium) when added to the oleyl amine capped Au nanoparticles suspended in toluene (a nonpolar medium), ligand exchange occurs because of the higher affinity of -SH compared to -NH$_2$ towards Au and results in a mixed ligand arrangement on the Au particles (see the molecular structure in Figure 3i). Here the amine bilayer formation is also possible.[9] Due to the increased polarity of the medium by addition of ethanol, the nonpolar part of MPTMS, that is, the silane end, tends to face the core with one of the amine ends of the bilayer of the oleyl amine facing the solvent as described above. Electron tomography performed on these structures shows that these are 3D spherical assemblies (see the movie in section S5) with a soft structure, therefore, attaining a dome-shaped morphology when deposited on a TEM grid, as shown by the 2D projection images in Figure 4a. Figure 4b presents the 3D rendering of the reconstructed tilt series showing the assembly of the particles. The orthoslices in Figure 4c further confirms that there is no SiO$_2$ matrix at the core supporting the Au particles assembly. These structures could coalesce to yield bigger assemblies in due course as in Figure 3b. Also, the free end of the MPTMS chain can undergo polymerization or self-assemble on the formed SiO$_2$ spheres depending on the MPTMS concentration.[14] In our case, initially the concentration is high enough

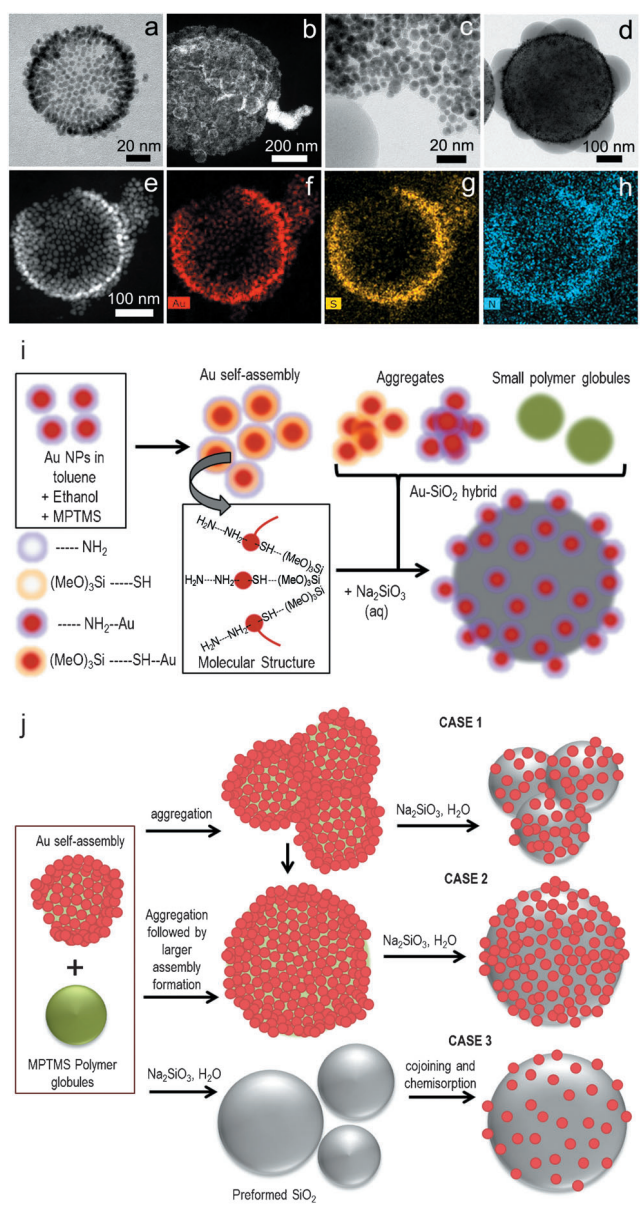

*Figure 3.* Intermediate stages of formation of the Au-SiO$_2$ hybrid as in a) self-assembled Au nanoparticles, b) aggregation of smaller self-assemblies, c) thiol- and amine-capped Au particles randomly aggregated and partially adsorbed on the preformed SiO$_2$ sphere, and d) conjoining of the smaller SiO$_2$ spheres to an Au-SiO$_2$ sphere. e–h) HAADF-STEM image of Au particle self-assembly and the elemental maps derived from Au-L, S-K, and N-K lines. Schematic representation describing i) the possible mechanistic route leading to the hybrid formation and j) possible routes to the Au-SiO$_2$ hybrid starting from the Au self-assemblies.

(0.56 m) to trigger polymerization and result in globular structures as shown in Figure 3i. Both of these soft-template structures could coalesce to yield bigger assemblies in due course; which can rearrange to produce larger hollow self-assemblies. However, when such assemblies are drop-cast on the grid we observe larger folded irregular shaped assemblies lying almost flat on the grid because of the soft template (see section S6). The silane end of the MPTMS can undergo thermal hydrolysis and subsequent condensation in the presence of an alkaline medium to form silica[15] but in the presence of sodium silicate, silica formation can be triggered even at room temperature.[8] Based on the above possible reactions, there can be three cases operating to result in an Au-SiO$_2$ hybrid (see Figure 3j). Case 1, in which the smaller assemblies cluster (as seen in Figure 3b) and the silane part facing the core of each assembly undergoes immediate hydrolysis and subsequent condensation to form SiO$_2$. However, this would result in an Au-particle-packed hybrid which is not the end product. In case 2, this cluster can spread out to yield a single larger hollow assembly, as described before, and further hydrolysis and condensation of the silane end would result in a hollow silica shell with particles on it which was also not the final product. Instead, with addition of Na$_2$SiO$_3$ (aq) solution, the reactant can diffuse in followed by SiO$_2$ nucleation and growth within the Au particles self-assemblies, forming a SiO$_2$ framework supporting the Au particles on the surface which is the final product. Another possibility, case 3, can be the growth of the preformed SiO$_2$ spheres by conjoining of the smaller ones and chemisorption of the Au assemblies on these by Si-O-Si linkage as in Figure 3c and d. This might result in a lower number of Au particles on the SiO$_2$ spheres as well as a nonuniform distribution. Thus, based on the final product observed we may conceive that both cases, 2 and 3, operate with case 2 dominating.

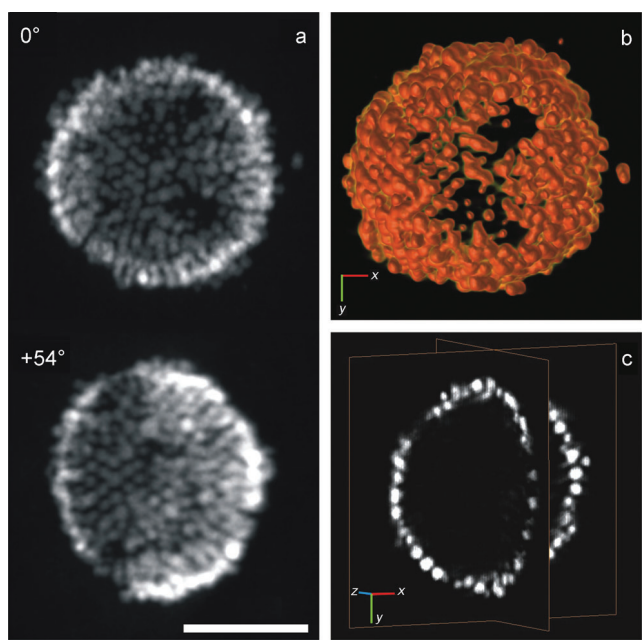

*Figure 4.* STEM tomography of the intermediate self-assembly of Au nanoparticles. a) 2D projection images at 0° and +54° tilt angles, indicating a dome-like shape (the scale bar is 50 nm). b) 3D volume rendering from the reconstruction of the aligned tilt series images, recorded by tilting the sample from −66° to +68° with 2° increment. c) Intersection of two different orthoslices showing that there are no particles located inside the assembly.

Heating experiments performed on the hybrid indicates interesting morphological changes. The nanostructures were initially heated up to 200 °C in a furnace in air and was further heated in situ up to 485 °C for microscopic study. HAADF-STEM imaging shows that the heterostructures did not undergo any morphological changes on heating (see Figure S7). The hybrid is stable up to 400 °C; however, heating at 485 °C shows few bigger Au particles. It is evident from the increased intensity in some regions because of the thickness contrast. From these 2D projection images we can conclude that there is no migration of the Au nanoparticles on the SiO$_2$ surface and hence no clustering or growth of the particles due to aggregation up to 400 °C. The clustering which appears, in 2D projection, after heating to 485 °C could also be apparent due to random migration of particles and their placement inside the matrix.[16] Electron tomography performed on the heated samples (200 °C, 485 °C) shows that the Au particles are indeed located inside the SiO$_2$ matrix, more at 485 °C, but not coarsened because of heating. It shows no change in the shape of the hybrid particles at 200 °C and 485 °C which is evident from the HAADF-STEM images, acquired at different tilt angles (see section S8). Figure 5a and d shows the volume rendering and the orthoslices in Figure 5b,e reveals the presence of particles on the surface as well as inside the spheres. It is clear from the orthoslices in Figure 5c,f that more Au particles migrated into the $SiO_2$ matrix (marked by arrows) at higher temperature (485°C; see the 3D animation in section S9). As there was no significant migration of the particles on the silica surface, the presence of the particles inside the spheres could be due to the collapse of the C-chain of the MPTMS ligands.[15]

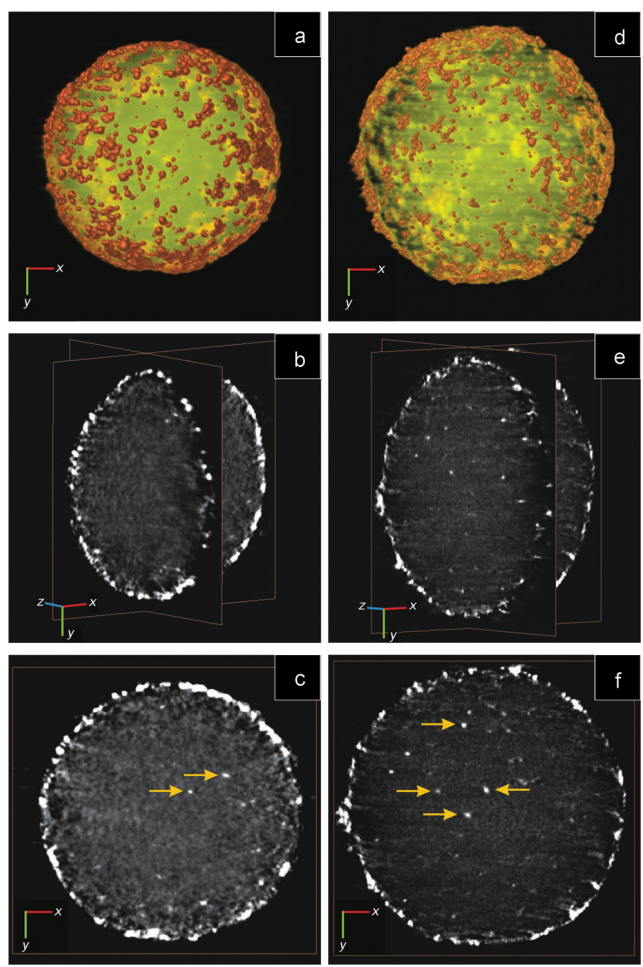

*Figure 5.* Electron tomography of an Au-SiO$_2$ particle heated to 200 and 485 °C. For 200 °C, a) 3D volume rendering obtained from the reconstruction of the aligned tilt series images, recorded by tilting the sample from −62° to +64° with 2° increments, showing that the Au particles are located at the surface of the sphere as well as inside the matrix. b) Intersection of two different orthoslices revealing that the particle remains spherical and c) orthoslice in the *xy*-plane confirming the presence of Au particles at the surface as well as inside the matrix of the sphere (as marked). d–f) The corresponding results for samples tilt from −66° to +70° with 2° increments at 485 °C. The orthoslices confirm the retention of the morphology of the hybrid and migration of a larger number of Au particles inside the matrix (as marked).

At sufficiently higher temperature (above 400°C), the damage of the organic mass is more which leads to migration of the particles inside the matrix in a random manner. Therefore, penetration of the Au particles into the matrix might result in finer channels leading to a porous structure. This study reveals that the polymeric $SiO_2$ matrix acts as a stable support for the Au particles, preventing them from surface migration and aggregation. The Au particles shape seems to remain intact at several places after heating, however, it is difficult to assess the shape of the particles that migrated inside the matrix (see Figure S10). We therefore conclude that the hybrid has a reasonable thermal stability and could be suitable as a catalyst for several reactions [17] or for any other related applications.

In summary, we demonstrated a novel chemical synthesis route to stabilize Au nanoparticles on polymeric $SiO_2$. A 3D characterization is carried out by electron tomography to understand the distribution of the Au particles, which has not been reported earlier, and elemental mapping confirms the composition in the hybrid. These techniques play a crucial role in understanding the shape, composition, and distribution of the heterounits of/in the hybrid, respectively. A mechanism is proposed based on the intermediate assembled structures of Au and $SiO_2$. A thermal stability test reveals the material to be a potential candidate for catalysis and biotechnology.